\documentclass[prl,twocolumn,showpacs]{revtex4}
\usepackage{graphicx}
\usepackage{bm}

\begin{document}
\title{Increasing entanglement between Gaussian states by coherent photon subtraction}

\author{Alexei Ourjoumtsev}
\author{Aur\'elien Dantan}
\author{Rosa Tualle-Brouri}
\author{Philippe Grangier}

\affiliation{Laboratoire Charles Fabry de l'Institut d'Optique, CNRS UMR 8501, 91403 Orsay, France}
\date{\today}
\email[Electronic address : alexei.ourjoumtsev@institutoptique.fr]{}

\begin{abstract}
We experimentally demonstrate that the entanglement between Gaussian entangled states can be increased by non-Gaussian operations. Coherent subtraction of single photons from Gaussian quadrature-entangled light pulses, created by a nondegenerate parametric amplifier, produces delocalized states with negative Wigner functions and complex structures more entangled than the initial states in terms of negativity. The experimental results are in very good agreement with the theoretical predictions.
\end{abstract}
\pacs{: 03.67.-a, 03.65.Wj, 42.50.Dv}
\maketitle

Entanglement plays a key role in quantum information processing (QIP). Entanglement distillation \cite{BennettPurification}, demonstrated for discrete-variable systems (ebits) in recent experiments \cite{GisinPurification,PanPurification,WaltherPurification}, allows one to produce strong entanglement between distant sites, initially sharing a larger set of weakly entangled states, and constitutes the basis of quantum repeaters, essential for long-distance quantum communications. An interesting alternative to discrete-level systems are quantum continuous variables (QCVs). In this case the information is encoded in the quadratures $\hat{x}$ and $\hat{p}$ of traveling light fields, which can be efficiently measured by homodyne detection. Optical parametric amplification allows one to produce quadrature-entangled beams, used in many QIP protocols. Together with linear optics, these tools preserve the Gaussian character of the states involved in most of QCV experiments~: the quasi-distributions (Wigner functions) of their quadratures remain Gaussian.
However, it has been shown that Gaussian entanglement distillation requires non-Gaussian operations \cite{EisertNoGaussPurif,JaromirNoGaussPurif,GiedkeNoGaussPurif}. Among several proposals \cite{DuanGaussPurif,EisertGaussPurif}, one of the simplest is the conditional subtraction of photons from Gaussian entangled beams \cite{KitagawaNonGaussEntangl,OpatrnyCondTeleport,CochraneCondTeleport,OlivaresCondTeleport}, by reflecting a small part of these beams towards two photon-counting avalanche photodiodes (APDs). If the reflectivity is low, a simultaneous detection of photons by the APDs heralds the subtraction of exactly one photon from each beam. Recently, such methods allowed the preparation and analysis of several states with negative Wigner functions, including one- and two-photon Fock states \cite{Lvovsky1photon,Zavatta1photon,Ourjoumtsev2photons}, delocalized single photons \cite{LvovskyEbit,BelliniTimeEbit} and  photon-subtracted squeezed states, very similar to quantum superpositions of coherent states with small amplitudes \cite{OurjoumtsevKittens,PolzikKittens}.

\begin{figure}[b]
\includegraphics[width=7cm]{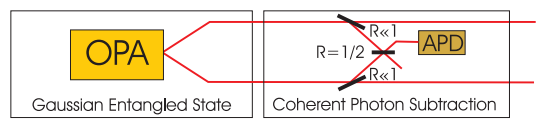}
\caption{Coherent photon subtraction from Gaussian entangled beams.}
\label{principe}
\end{figure}

In this Letter, we experimentally demonstrate that non-Gaussian operations allow us to increase the entanglement between Gaussian states, with a  protocol presented on Fig.~\ref{principe}. An optical parametric amplifier (OPA) produces Gaussian quadrature-entangled light pulses, known as two-mode squeezed states \cite{PulsedEPR}. We pick off small fractions of these beams, which interfere with a well-defined phase on a 50/50 beam splitter (BS), and we detect photons in one of the BS outputs. This way, we subtract a \textit{single} photon delocalized in the two beams and prepare a complex quantum state with a negative two-mode Wigner function. We determine a range of experimental parameters where the entanglement of the prepared state, quantified by the negativity \cite{WernerNegativity}, is significantly higher compared to that of the initial Gaussian state.

Operating with single photon counts rather than with coincidences as proposed e. g. in \cite{KitagawaNonGaussEntangl,OpatrnyCondTeleport,CochraneCondTeleport,OlivaresCondTeleport,BellCerf,BellCar}, this protocol allows for much higher generation rates and produces states more robust to experimental imperfections (see below). Besides, it is more efficient at moderate OPA gain: in the zero-gain limit, the detection of a photon transforms a state with almost no entanglement into a maximally entangled ebit state $(\vert 1 0 \rangle + \vert 0 1 \rangle)/\sqrt{2}$. With the higher gain (up to 3 dB) used in the present experiment, the generated states have a much richer structure as is shown below.

\begin{figure}[t]
\includegraphics[width=7cm]{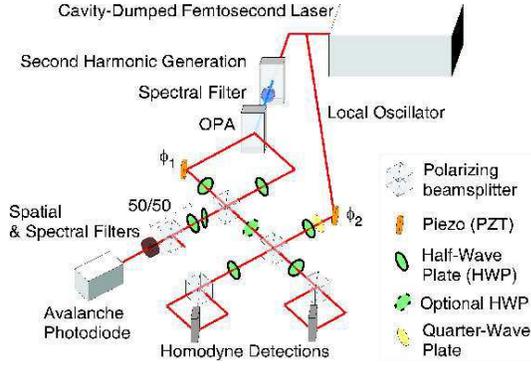}
\caption{Experimental setup.}
\label{setup}
\end{figure}

Our experimental setup is presented in Fig. \ref{setup}.
Nearly Fourier-limited femtosecond pulses (180 fs, 40 nJ), produced by a Ti:sapphire laser with a 800 kHz repetition rate, are frequency doubled by a single pass in a $100 \; \mu m$-thick
type I noncritically phase-matched potassium niobate (KNbO$_{3}$) crystal. The frequency-doubled beam pumps an
identical crystal used as an optical parametric amplifier (OPA), generating Gaussian quadrature-entangled pulses spatially separated by an angle of $10^{\circ}$. Adjusting the pump power allows us to vary the two-mode squeezing between 0 and 3.5 dB. The photon pickoff beam splitters are realized with a single polarizing beam splitter (PBS) cube, where the signal and idler beams are recombined spatially but remain separated in polarization. A small adjustable fraction $R$ of both beams is sent into the APD channel, where they interfere on a 50/50 BS. A tilted half-wave plate compensates for residual birefringence. An APD detects one of the 50/50 BS outputs after spatial and spectral filtering. The signal and idler beams transmitted through the pickoff beam splitter are projected into a non-Gaussian state by an APD detection. They are spatially separated on another PBS, where they are combined with bright local oscillator beams. A quarter-wave and a half-wave plate allow us to prepare two local oscillators with equal intensities and a well-defined relative phase. The signal and idler beams are analyzed by two time-resolved homodyne detections, which sample each individual pulse, measuring one quadrature $x_{1,2}(\theta_{1,2})$ in phase with the local oscillator.

In this setup, all the relative phases except $\phi_1$ and $\phi_2$ (see Fig. \ref{setup})
are precisely adjusted with wave plates. Phase fluctuations concern only the initial two-mode squeezed state, where the phase difference is not defined and plays no role, and the slow (thermal and acoustic) phase sum fluctuations simply rotate the two-mode squeezing ellipse and can be compensated by shifting the common phase $\phi_2$ of the local oscillators. This phase can be scanned with a piezo translator, and rapidly measured using the unconditioned two-mode squeezing variance.

Quantum states with negative Wigner functions are very sensitive to experimental imperfections. In our case, the most important issue are spurious APD trigger events, due to imperfect filtering, limited qualities of the optical beams, imperfect mode-matching between the subtracted beams, and APD dark counts. An APD count corresponds to the desired subtraction event with a success probability $\xi<1$. This explains why single-photon protocols are more robust than two-photon ones, where the total success probability is only $\xi^2$. Another issue is the OPA excess noise. To describe it, we can consider that a first amplification process creates a pure entangled state with a two-mode squeezing variance $s=e^{-2r}$, and that each of the resulting modes is independently amplified with a gain $h=\cosh^2(\gamma r)$ by a phase-independent amplifier
 with a relative efficiency $\gamma$. The finite homodyne efficiency $\eta$ and the homodyne excess noise $e$ also deteriorate the measured data. However, they are not involved in the generation process but only in the analysis, and we can correct for their effects in order to determine the actual Wigner function of the generated state. Even with none of these imperfections, this protocol would still be limited by the finite pick-off BS reflectivity $R$, required for a sufficient APD count rate but inducing losses on the transmitted beam. The limited overall efficiency $\mu=5\%$ of the APD channel has little effect in this experiment.

A detailed analytic model \cite{OurjoumtsevKittens,Ourjoumtsev2photons} including all these imperfections yields an expression for the Wigner function W of the state studied in our experiment~:
\begin{equation}
W(x_1,p_1,x_2,p_2) = W_{s}(x_+,p_+) W_{c}(x_-,p_-)
\label{eqnWigner}
\end{equation}
where $x_{\pm} = \frac{x_1 \pm
x_2}{\sqrt{2}}$, $p_{\pm} = \frac{p_1 \pm
p_2}{\sqrt{2}}$, $W_s$ is the Wigner function of a single-mode squeezed state, and $W_c$
corresponds to a photon-subtracted squeezed state analyzed in \cite{OurjoumtsevKittens}. More explicitly :
\begin{eqnarray}
\nonumber
W_{s}(x,p) &=& \exp\left(-x^2/a-p^2/b \right)/(\pi \sqrt{ab})\\
\nonumber
W_{c}(x,p) &=& W_{s}(x,p)\left[ \frac{2A}{a^2} x^2 + \frac{2B}{b^2} p^2 +1-\frac{A}{a}-\frac{B}{b}\right]\\
\nonumber
a(s) &=& b(1/s) \;=\; 1+e+\eta (1-R)(h s + h -2)\\
\nonumber
A(s) &=& B(1/s) \;=\; \frac{\eta \; \xi \; (1-R)(h s + h -2)^2}{h(s+1/s) + 2h -4}
\end{eqnarray}
In this experiment the photon-subtracted state is ``delocalized" into two spatially separated modes 1 and 2 and revealed by measuring the correlations between identical quadratures, the anticorrelations remaining in the initial squeezed state.

\begin{figure}
\includegraphics[height=2.8cm]{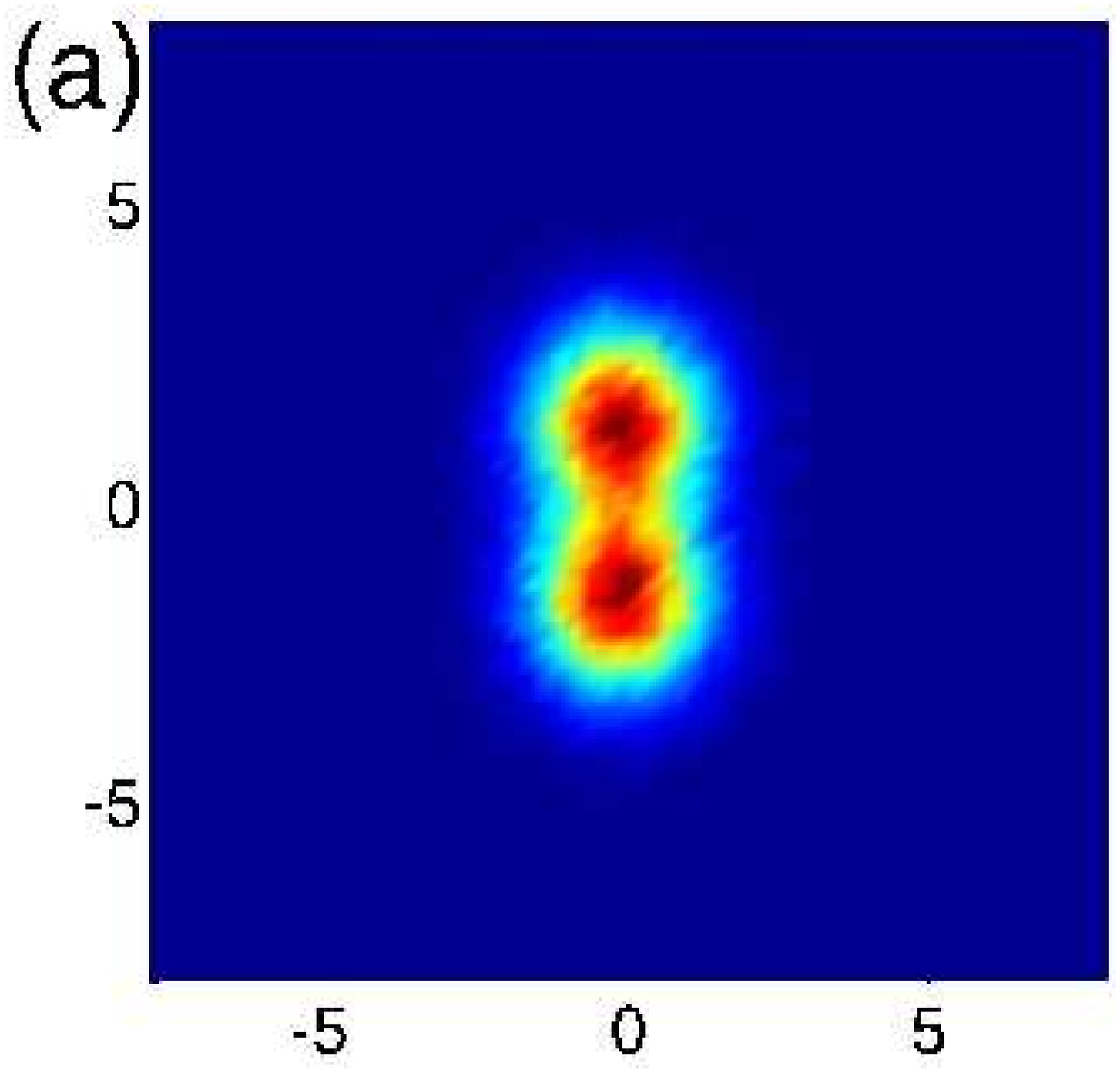}
\includegraphics[height=2.8cm]{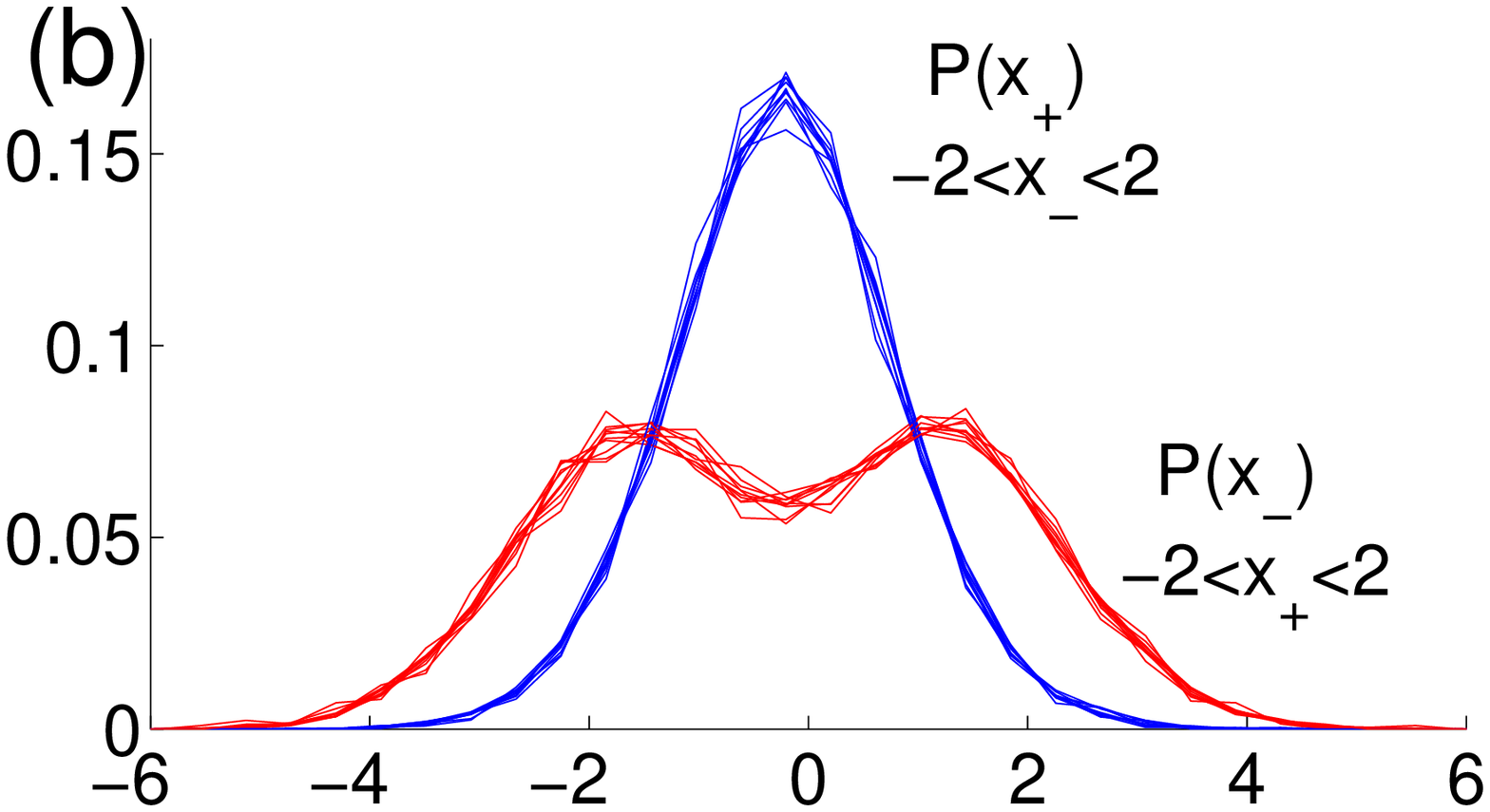}
\caption{State separability test after interference between signal
and idler beams : (a) Joint distribution $P(x_+,x_-)$, (b)
Distributions $P_s(x_+)$ and $P_c(x_-)$, for 11 values of  $x_-$ (resp. $x_+$),  chosen between -2 and +2. This separability was verified for several randomly
chosen values of $\theta_+$ and $\theta_-$ (here $\theta_+=20^\circ$
and $\theta_-=50^\circ$).}
 \label{separability}
\end{figure}

Without assuming any particular shape for $W_s$ and $W_c$, we can
experimentally show that the state becomes separable if we make a
joint measurement, transforming $x_{1,2}$ into $x_{\pm}$ by rotating the polarizations by $45^\circ$ with the
optional half-wave plate shown on Fig. \ref{setup}. We then observe that the
quadratures measured by one detection do not depend on the other
(see Fig. \ref{separability}). For every $\theta_\pm$ the joint
distribution, and hence the Wigner function, becomes factorable~:
$P(x_+(\theta_+),x_-(\theta_-))=P_s(x_+(\theta_+))P_c(x_-(\theta_-))$.
It means that one can fix $\theta_-=\theta_+=\theta$ and scan
$\theta$ to perform a complete tomography of this state.

This property considerably simplifies the experimental analysis. We perform direct homodyne measurements of the entangled quadratures $x_{1,2}( \theta)$, keeping the entangled modes 1 and 2 separated without mixing them. We use the fact that the state factorizes in the $x_+,x_-$ basis to reconstruct it from a limited set of data : instead of a time-consuming general two-mode tomography, which requires to measure $x_1(\theta_1),x_2(\theta_2)$ with all possible combinations of phases, we can restrict ourselves to $\theta_1=\theta_2=\theta$.
In practice, we set the relative phase between the local oscillators
to zero, scan the common phase $\phi_2=\theta$, measure
$x_{1,2}(\theta)$ and calculate $x_{+,-}(\theta)$. We reconstruct
the distributions $P_c(x_-(\theta))$ and $P_s(x_+(\theta))$ for
several phases. We observe that the measured distributions are
invariant under  $\theta\rightarrow\pi\pm\theta$, so we restrict
the analysis to $0\leq\theta\leq\pi/2$. Typically, we measure 6 to
12 different quadrature distributions, with 10000 to 20000 data
points each. A numeric Radon transform allows us to reconstruct the
uncorrected Wigner functions $W_c$ and $W_s$. We can correct for the
homodyne detection losses ($\eta=70\%$, $e=1\%$ of the shot noise)
using a maximal-likelihood algorithm
\cite{RehacekMaxLik,LvovskyMaxLik} to obtain the Wigner function $W$
of the generated state. We use $W$ to calculate the density matrix $\rho$ of the state and obtain
its entanglement, given by the negativity
$\mathcal{N}=\frac{\parallel \rho^{T_1} \parallel_1 -1}{2}$, where
$T_1$ is the partial transposition operation
\cite{WernerNegativity}.

\begin{figure}
\includegraphics[height=3.5cm,width=3.5cm]{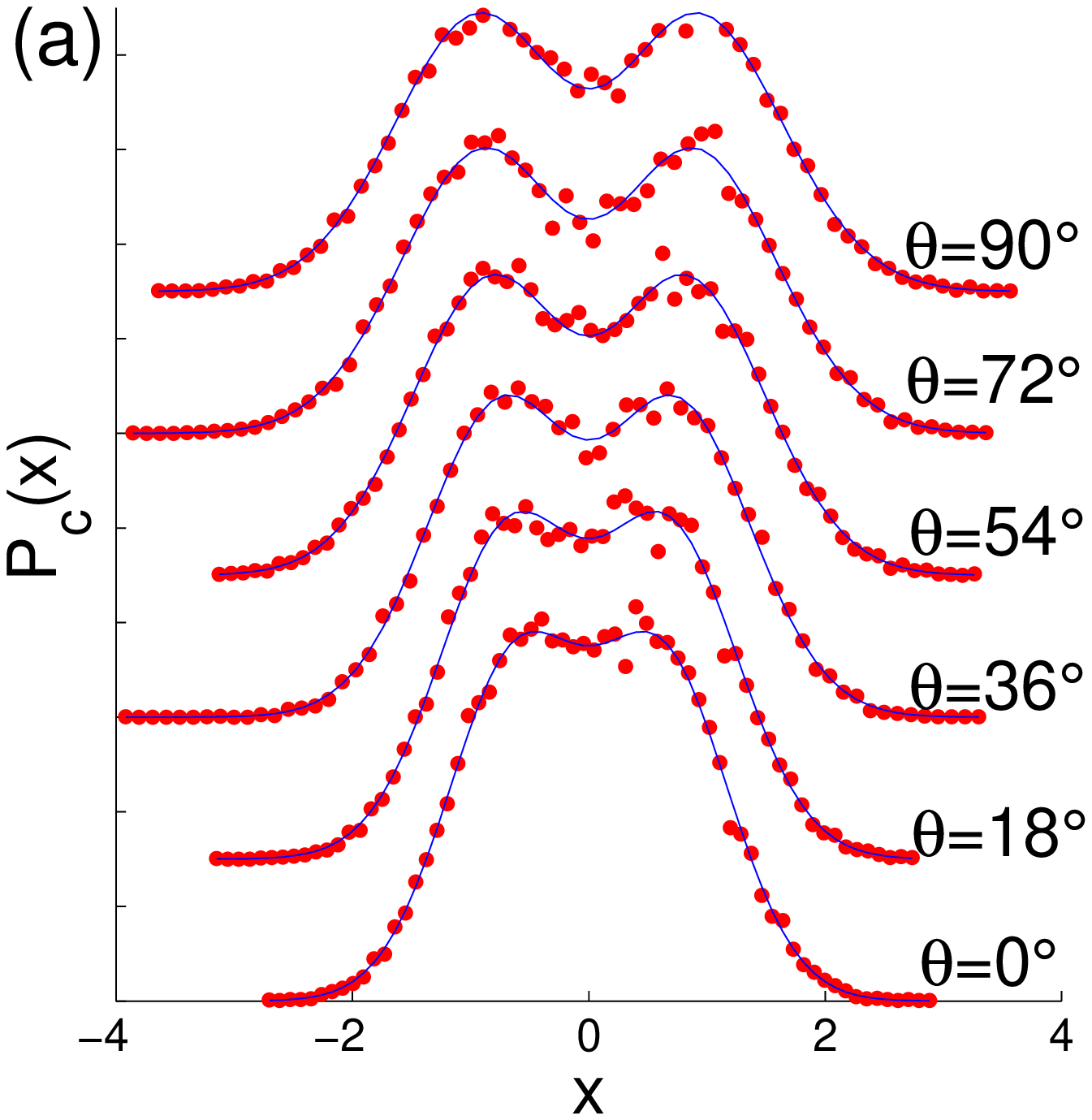}
\includegraphics[height=3.5cm,width=5cm]{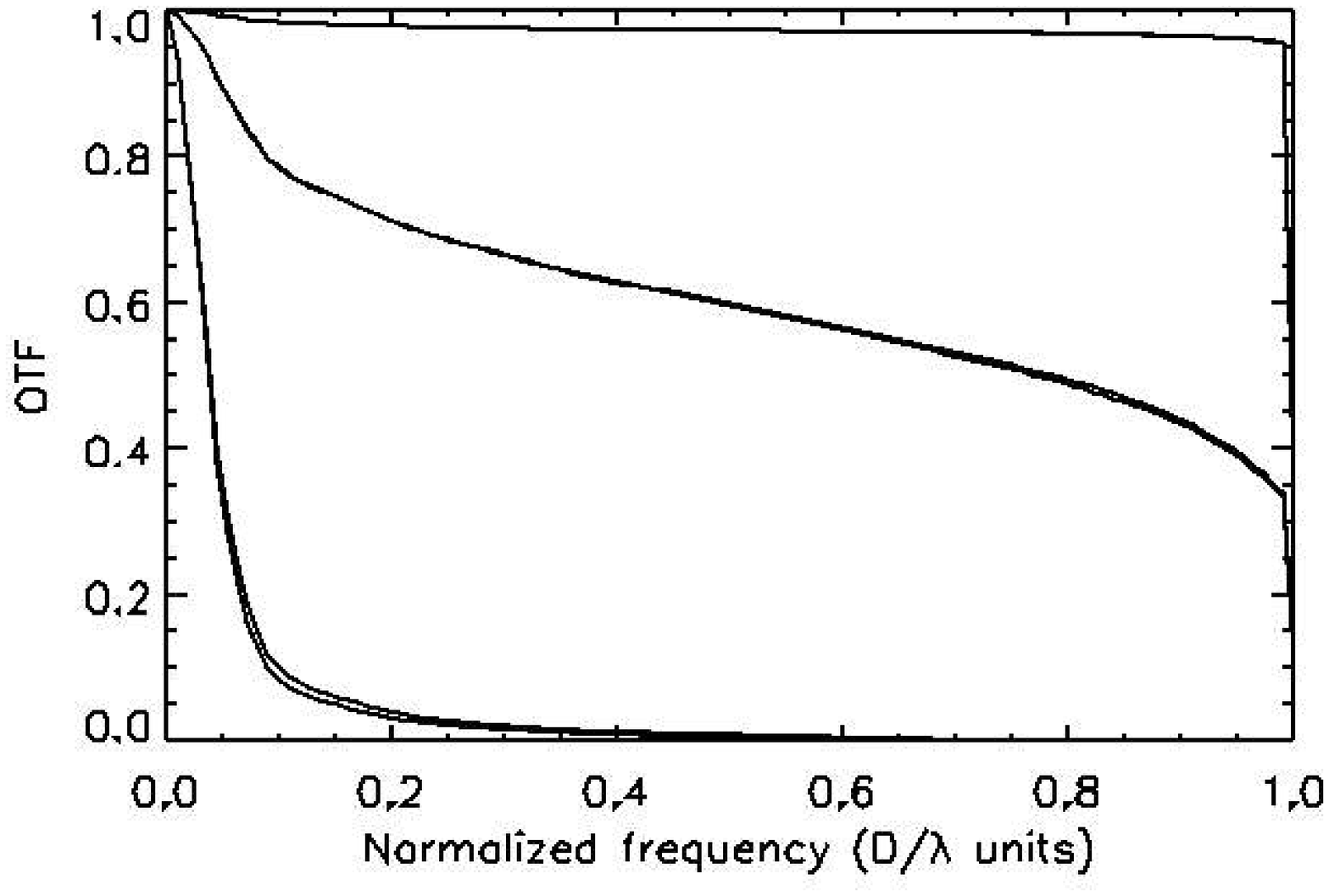}

\includegraphics[height=3.5cm,width=3.5cm]{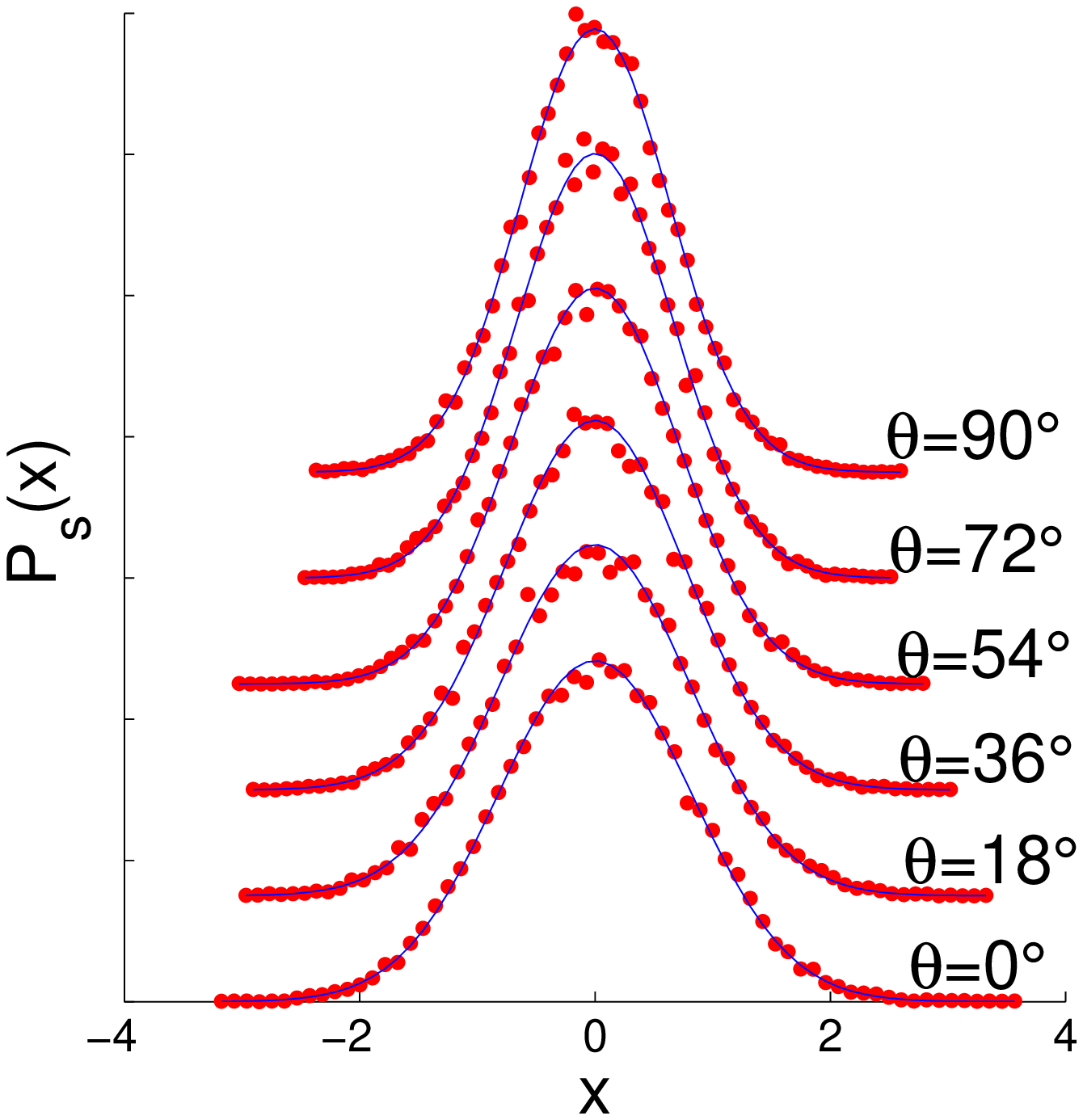}
\includegraphics[height=3.5cm,width=5cm]{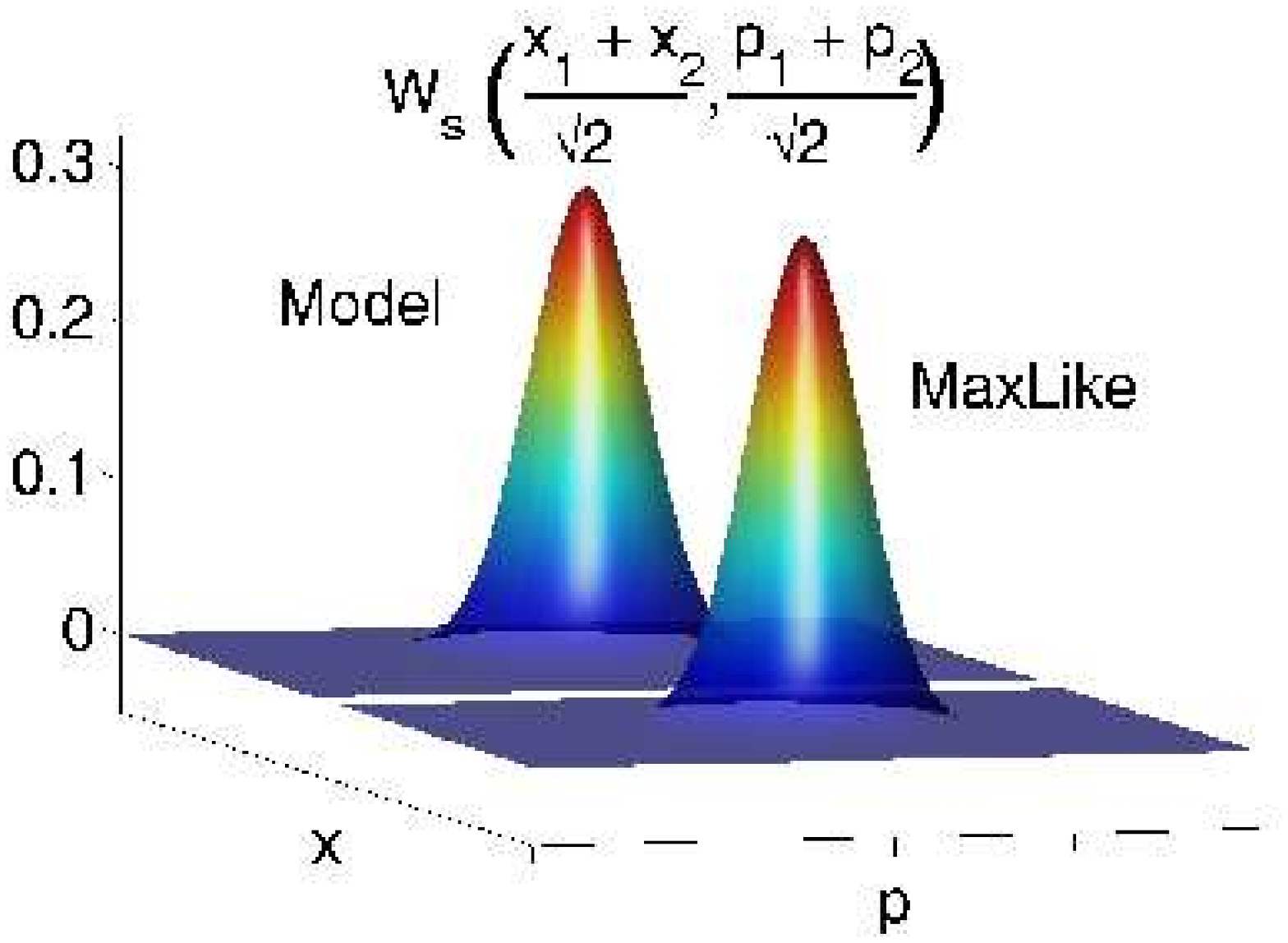}

\caption{(a) Set of experimentally measured quadrature distributions
(dots), compared to those reconstructed from our model (solid line).
(b) Wigner function corrected for homodyne detection losses,
obtained with a standard maximal-mikelihood algorithm (MaxLike),
compared to the result of our model. This state is produced with 1.8
dB of squeezing and $R=5\%$.} \label{ModelMaxLike}
\end{figure}

Figure \ref{ModelMaxLike} presents the tomography of a state
produced with 1.8 dB of squeezing and a BS reflectivity $R=5\%$. The
Wigner function, corrected for detection losses, is clearly negative
: $W_c(0)=-0.13 \pm 0.01$ ($0.01 \pm 0.01$ before correction). The
entanglement of this state is $\mathcal{N}=0.34 \pm 0.02$, whereas
for the initial state (before the pickoff BS) $\mathcal{N}_0=0.24
\pm 0.01$.

In Refs. \cite{OurjoumtsevKittens,Ourjoumtsev2photons} we
demonstrated another analysis method, more constrained but also much
faster and closer to the physics of the experiment. If we assume
that the Wigner function has the form defined in Eq.
\ref{eqnWigner}, we can easily extract the parameters $a$, $A$, $b$,
$B$ from the second and fourth moments of the measured
distributions, and determine the Wigner function, the density matrix, and
the quadrature distributions of the measured state. For a given
squeezing, one can also obtain from $a$, $A$, $b$ and $B$ the values
of all the experimental parameters introduced above. To
correct for homodyne losses, we simply calculate the Wigner
function that we would measure with an ideal detection ($\eta=1$ and
$e=0$), using the values extracted from the experimental data for
all the other parameters. As shown in Fig. \ref{ModelMaxLike}, the
distributions reconstructed with this method are in excellent
agreement with those directly extracted from the data, and the
Wigner function is almost indistinguishable
from the one obtained with the maximal-likelihood algorithm. Both
methods give the same values for the negativity.

\begin{figure}[b]
\includegraphics[width=2.8cm]{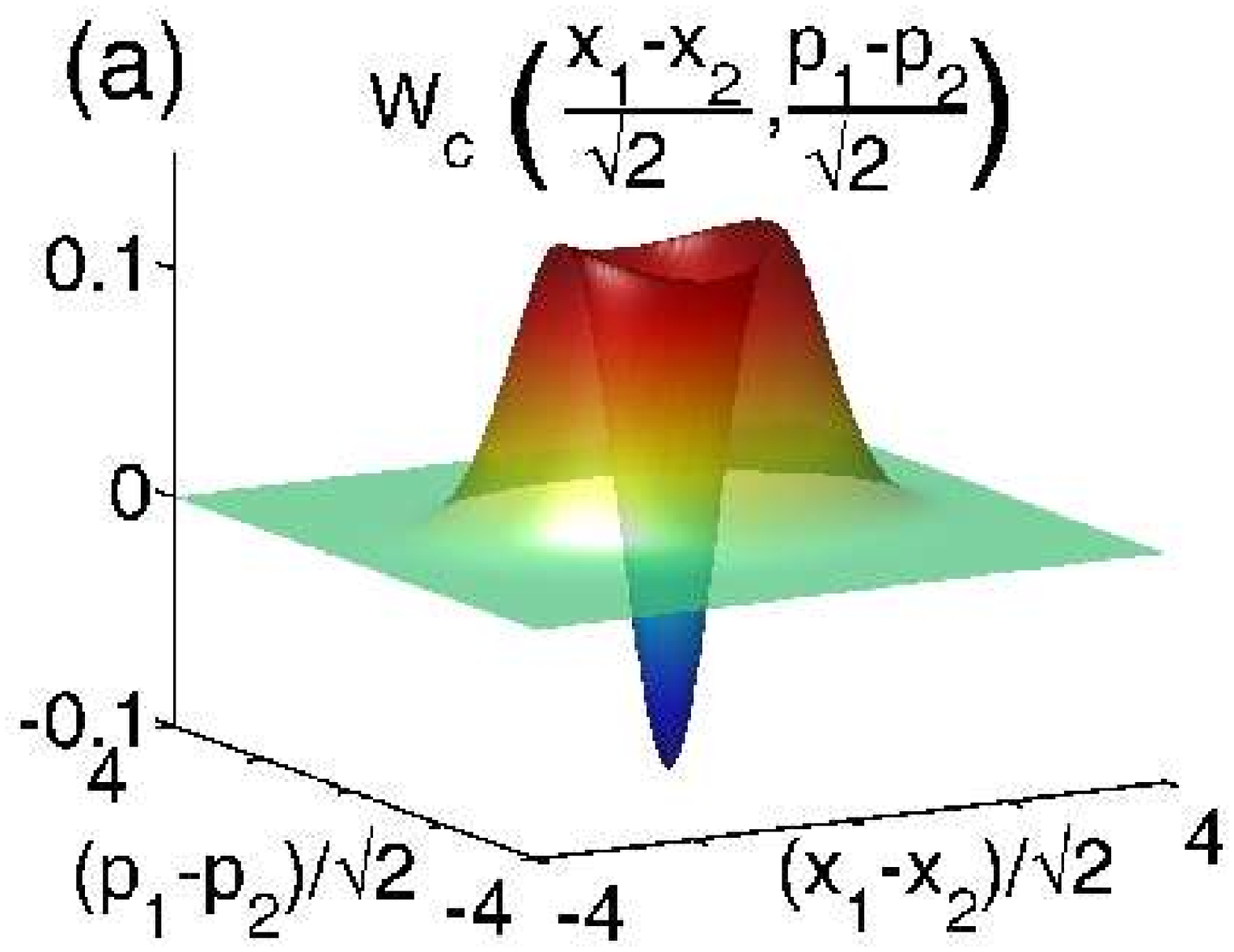}
\includegraphics[width=2.8cm]{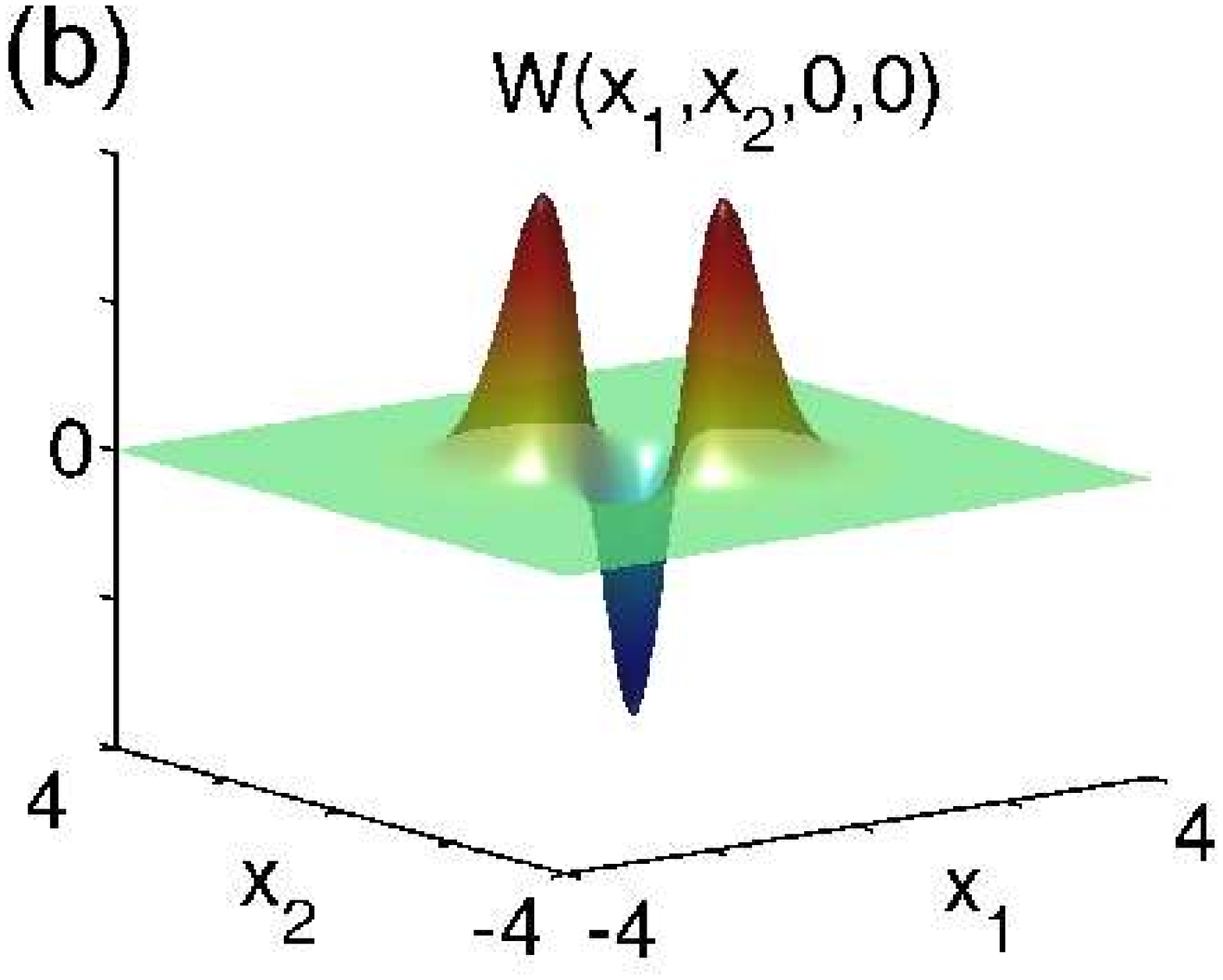}
\includegraphics[width=2.8cm]{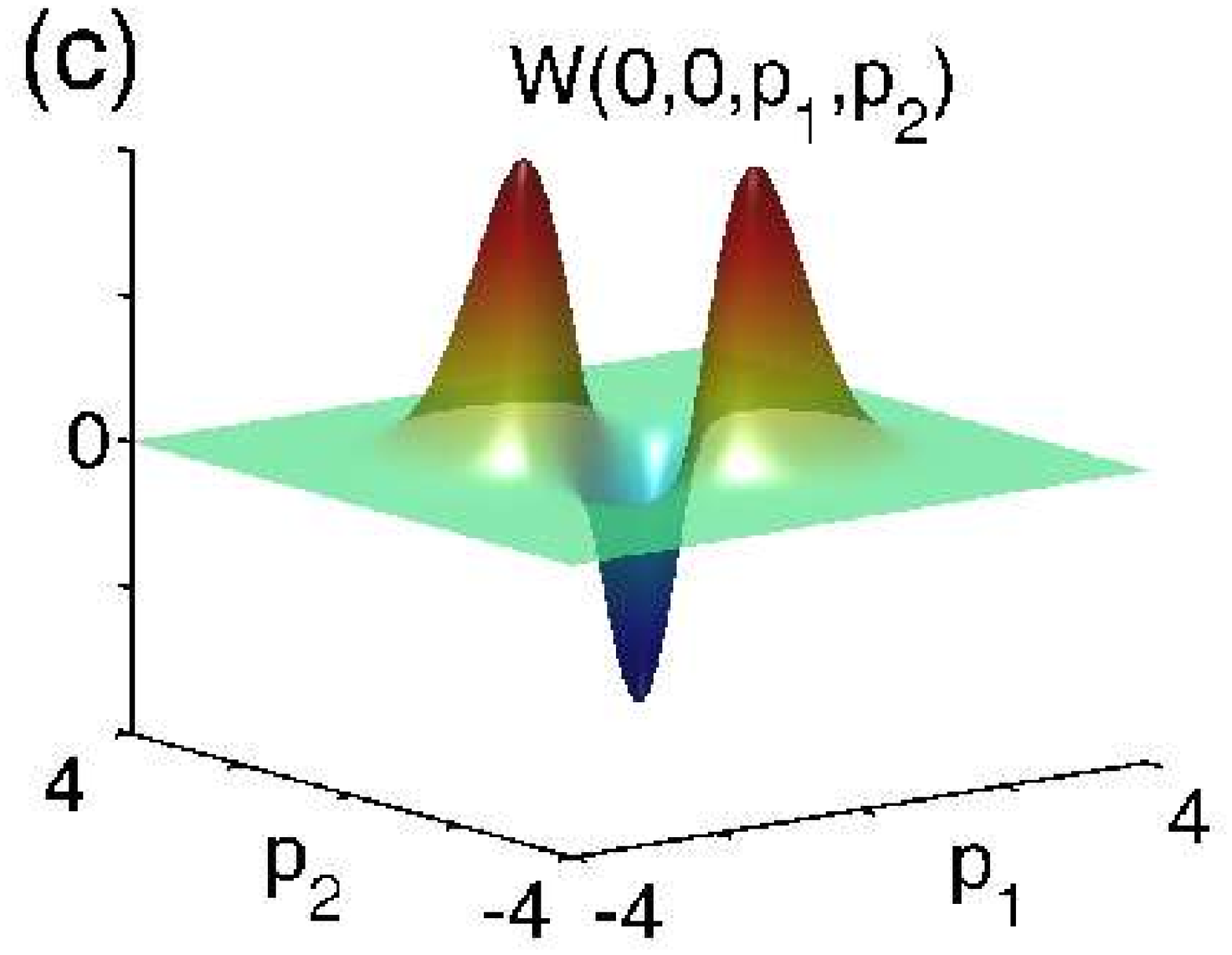}

\includegraphics[width=2.8cm]{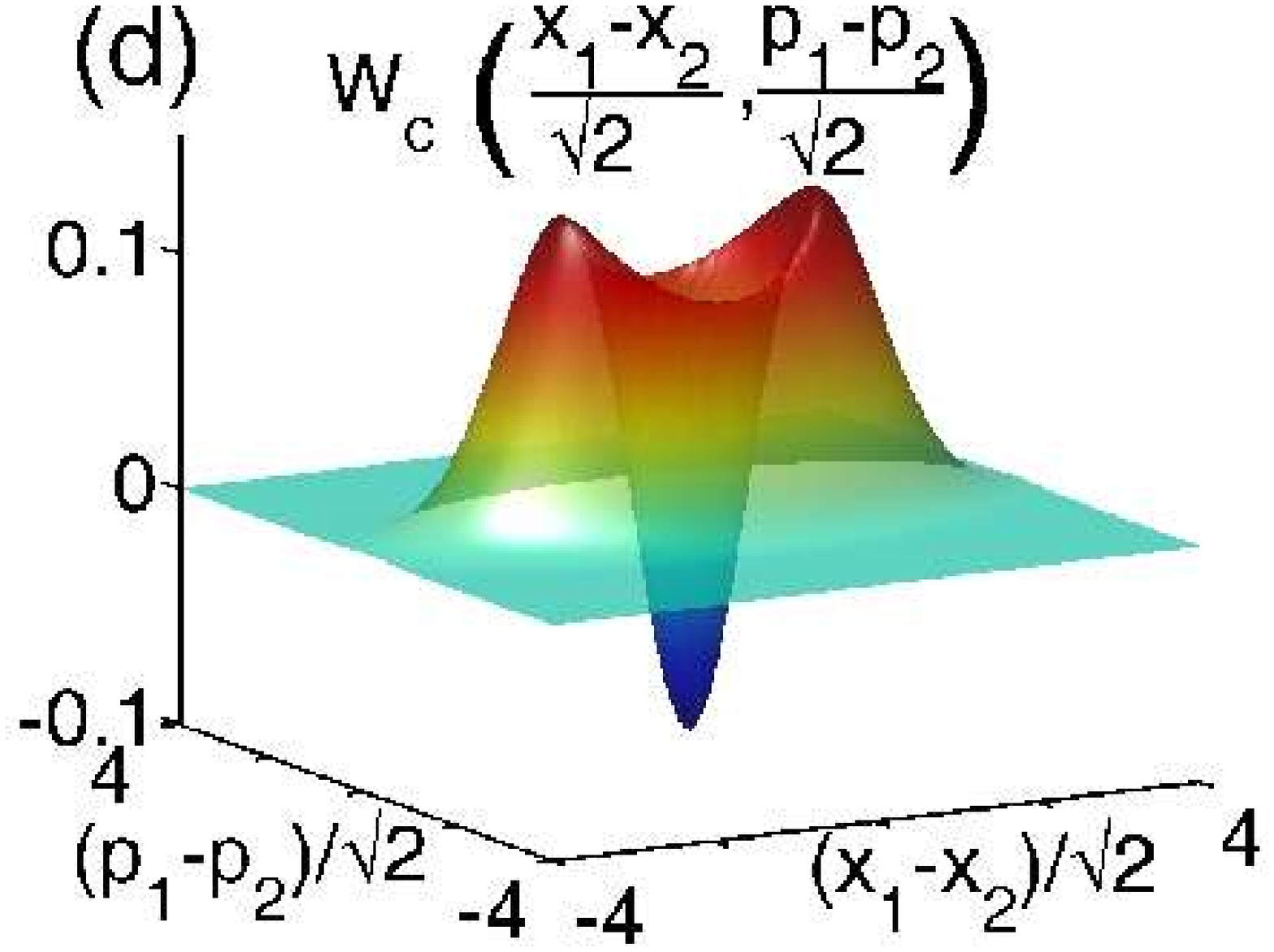}
\includegraphics[width=2.8cm]{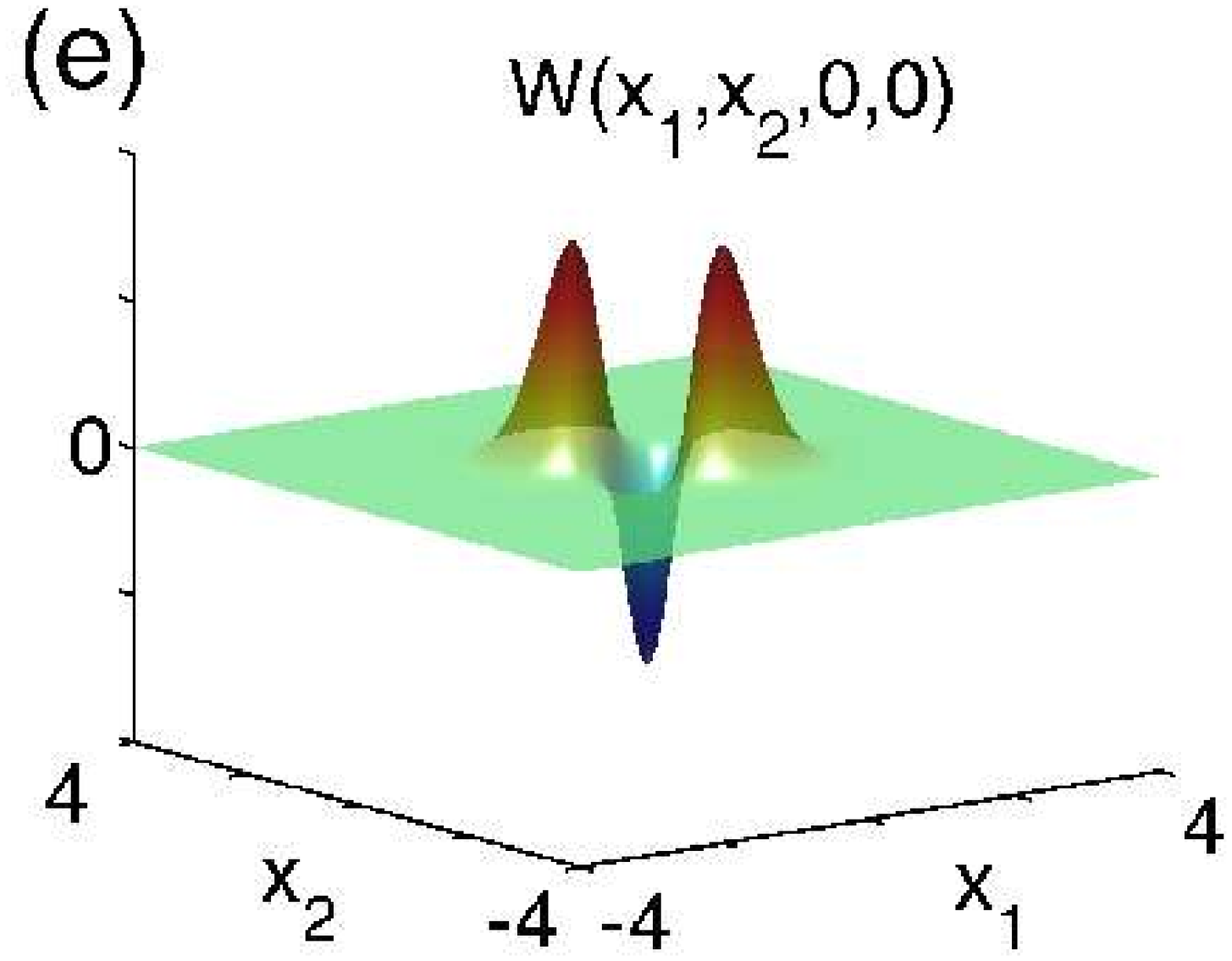}
\includegraphics[width=2.8cm]{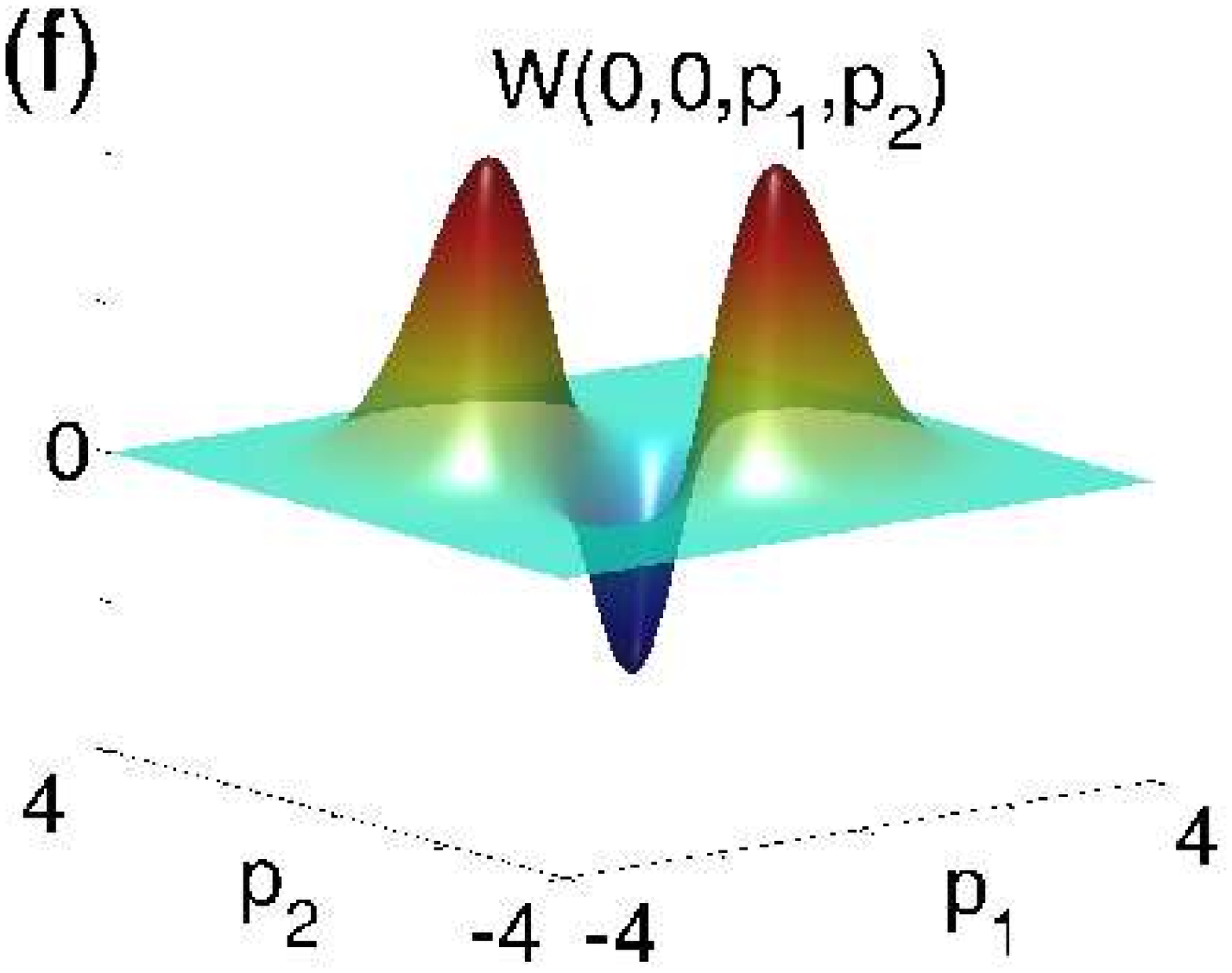}
\caption{Tomography of two states produced with 1.3 dB (first row) and
3.2 dB (second row) of squeezing and $R=10\%$: three different
cuts of the two-mode Wigner function.} \label{tomographies}
\end{figure}

A natural question to ask is whether this protocol works for an
arbitrary squeezing or if, when the initial state is already
strongly entangled, by performing an imperfect photon subtraction we actually lose more entanglement than we gain. It has already been
shown that, when the pick-off beam splitters have a finite
reflectivity, subtracting one photon from each of the Gaussian
entangled beams may actually decrease the entanglement
\cite{KitagawaNonGaussEntangl}. Using our model, we can take into
account all the other experimental parameters to derive an analytic
expression for $\rho^{T_1}$, which can be
diagonalized numerically in a few seconds to obtain the expected
negativity of a given state. We found that the experimental
imperfections have a very strong effect. For example, for an initial
squeezing of 3 dB, the negativity increases ideally from
$\mathcal{N}_0=0.50$ to $\mathcal{N}=0.90$. If we assume $R=3\%$ for
the pick-off BS, then $\mathcal{N}=0.81$, but if we include the
average values of experimental parameters involved in the state
preparation, $\langle \gamma \rangle = 0.22$ and $\langle \xi
\rangle=0.78$, $\mathcal{N}$ drops down to 0.51, whereas for the
initial state, only slightly affected by $\gamma$,
$\mathcal{N}_0=0.49$.

To verify this experimentally, we performed several tomographies for
different BS reflectivities and degrees of squeezing. Figure
\ref{tomographies} presents two tomographies with
$R=10\%$ for a small (1.3 dB) and a high (3.2 dB) squeezing. As
expected,
the Wigner function becomes more phase-dependent and less negative as the state becomes
``bigger" and more sensitive to decoherence.

\begin{figure}[t]
\includegraphics[width=7cm]{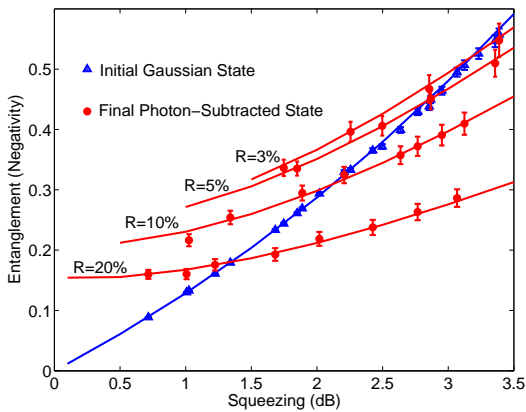}
\caption{Entanglement negativity of the initial and final states as
a function of squeezing for several pickoff BS reflectivities,
corrected for homodyne detection losses. Solid lines are theoretical
calculations using the average values of the experimental
parameters.} \label{negativities}
\end{figure}

Figure \ref{negativities} shows the entanglement negativity of the
photon-subtracted states and the corresponding initial states.
The solid lines are theoretical calculations using the average values of
the experimental parameters involved in the state preparation,
$\langle \gamma \rangle = 0.22$ and $\langle \xi \rangle=0.78$.
Two
domains appear on the graph: the upper left, where this
process actually increases the entanglement,
and the lower right, where the initial state is too sensitive
to the added losses.
As expected, this protocol is particularly
efficient at low squeezing. One can show that when the squeezing
and hence the entanglement of the initial state tend to $0$, the
negativity of the final state has a nonzero limit
$\mathcal{N}_{r\rightarrow 0}=\frac{\sqrt{C^2+(1-C)^2}-(1-C)}{2}$
where $C=\frac{\xi (1-R)}{1+\gamma^2}$. At low squeezing, experimental imperfections have a moderate effect
on the state. We nevertheless succeed in improving the
negativity of a state with up to $3$ dB of squeezing, which
corresponds to a strong squeezing regime where small
experimental improvements strongly affect
the performance of the protocol. For example, increasing
$\xi$ by a mere $4\%$ with $R=3\%$ should displace the crossover
point from 3 to 4 dB (in other experiments, where mode matching
between subtracted photons was not an issue, $\xi$ reached $0.9$
\cite{Ourjoumtsev2photons}).

In conclusion, the present photon subtraction protocol allows one to
increase the entanglement between Gaussian states with up to 3 dB of
squeezing, and even small experimental improvements should
significantly increase this limit. For QIP protocols specifically
requiring Gaussian entanglement, these non-Gaussian states could in
principle be used as a starting point for a ``Gaussification"
procedure \cite{EisertGaussPurif}. This demonstrates one of the key
steps required for long distance quantum communications with
continuous variables.

\begin{acknowledgments}
This work is supported by the EU IST/FET project COVAQIAL, and by
the French ANR/PNANO  project IRCOQ.
\end{acknowledgments}

\bibliography{OurjoumtGaussEnt}

\end{document}